\documentclass[conference]{IEEEtran}
\usepackage{latexsym}
\usepackage{graphicx}
\usepackage{amsfonts,amssymb,amsmath}
\usepackage{algorithm,algorithmic}
\usepackage{hyperref}

\usepackage[T1]{fontenc}
\usepackage{cite}
\usepackage{subcaption}
\usepackage{comment}
\usepackage{amsthm}
\usepackage{overpic}
\usepackage{steinmetz}
\usepackage{array}
\usepackage{url}
\usepackage{xcolor}
\IEEEoverridecommandlockouts

\theoremstyle{plain}
\newtheorem{theorem}{Theorem}

\newcommand{\argmax}[1]{{\underset{{#1}}{\mathrm{arg\,max}}}}
\newcommand{\argmin}[1]{{\underset{{#1}}{\mathrm{arg\,min}}}}
\newcommand{\vect}[1]{\mathbf{#1}}

\def\diag{\mathrm{diag}}

\def\Htran{\mbox{\tiny $\mathrm{H}$}}
\def\Ttran{\mbox{\tiny $\mathrm{T}$}}
\def\CN{\mathcal{N}_{\mathbb{C}}} 

\begin{document}

\title{Efficient LOS Channel Estimation for RIS-Aided Communications Under Non-Stationary Mobility}

\author{Mehdi Haghshenas\textsuperscript{1}, Parisa Ramezani\textsuperscript{2}, Emil Bj{\"o}rnson\textsuperscript{2} \\
\IEEEauthorblockA{\textit{\textsuperscript{1} Department of Electronics, Information and Bioengineering, Politecnico di Milano, 20133 Milan, Italy} \\ \textit{\textsuperscript{2}Department of Computer Science, KTH Royal Institute of Technology, SE-100 44 Stockholm, Sweden}\\ Email: mehdi.haghshenas@polimi.it, \! \{parram, emilbjo\}@kth.se}%
\thanks{This work was supported by the FFL18-0277 grant from SSF.}
}

\maketitle

\begin{abstract}
Reconfigurable intelligent surface (RIS) is a newly-emerged technology that, with its unique features, is considered to be a game changer for future wireless networks. Channel estimation is one of the most critical challenges for the realization of RIS-assisted communications. Non-parametric channel estimation techniques are inefficient due to the huge pilot dimensionality that stems from the large number of RIS elements. The challenge becomes more serious if we consider the mobility of the users where the channel needs to be re-estimated whenever the user moves to a new location. This paper develops a novel maximum likelihood estimator (MLE) for jointly estimating the line-of-sight (LOS) channel from the user to the RIS and the direct channel between the user and the base station. By smartly refining the RIS configuration during the channel estimation procedure, we show that the channels can be accurately estimated with only a few pilot transmissions---much fewer than the number of RIS elements. The proposed scheme is also shown to be capable of effectively tracking the channel when the user moves around in a continuous but non-stationary manner with varying LOS angles.%
\end{abstract}
\begin{IEEEkeywords}
Reconfigurable intelligent surface, parametric channel estimation, maximum likelihood estimator.
\end{IEEEkeywords}

\maketitle

\section{Introduction}

A reconfigurable intelligent surface (RIS) can shape the propagation environment between a wireless transmitter and receiver \cite{Huang2018a, Renzo2020b,Bjornson2020a}.
The prevalent use case is to deploy an RIS within line-of-sight (LOS) from a base station (BS) and then configure it to reflect the BS’s signals toward a user that has LOS to the RIS but not to the BS \cite{Bjornson2022a}. This is because the RIS cannot manage the highly frequency-selective fading in non-LOS scenarios and hardly beats a direct LOS path. 
Channel state information (CSI) is essential for RIS configuration and can be acquired through pilot signaling. Unfortunately, the pilot length required by the least-squares (LS) estimator \cite{Jensen2020a} and minimum mean-squared error estimator \cite{Wang2020a} is proportional to the number of RIS elements, which can be in the order of hundreds.
This massive pilot requirement makes it essential to find an answer to the question posed in \cite{Bjornson2020a}: ``\emph{Can an RIS be real-time reconfigured to manage user mobility?}’’

Dynamic RIS reconfiguration under user mobility has been considered in several recent works, with a focus on scenarios with stationary fading processes.
Jakes’ model was used in \cite{Papazafeiropoulos2022a,Zhang2022a} to describe the temporal fading correlation over non-LOS channels with Rayleigh fading and known statistics.
LOS channels were treated in \cite{Xu2022,Chen2022a} using Rician fading models with similar temporal correlation and LOS paths that are constant over time except for the phase shifts.
These approaches can track the channel aging for users moving in a tiny region where the channel statistics can be modeled as stationary but cannot manage large-scale mobility (e.g., vehicles or people passing by an RIS).
In LOS scenarios, where the optimal RIS configuration reflects the signal via the LOS path \cite{Bjornson2022a}, the RIS must only be reconfigured when the user moves out of the main lobe of the reflected signal; that is, when the channel statistics are changing so the prior methods are insufficient.

Pilots must be transmitted more frequently under mobility, thus it is problematic to use the approaches in \cite{Papazafeiropoulos2022a,Zhang2022a,Xu2022,Chen2022a} that require a pilot length equal to the number of RIS elements (plus one for the direct path).
There are ways to utilize the channel structure to reduce the pilot length. One approach is to treat adjacent RIS elements as a single element \cite{CYou2020}, at the cost of reduced beamforming gain. By limiting the scope to a LOS scenario, one can exploit the resulting channel structure to estimate the complete channels using shorter pilots, using compressive sensing or array signal processing methods \cite{Swindlehurst2022a}. 
For example, parametric maximum likelihood estimators (MLEs) were derived in \cite{Wang2021a,Bjornson2022b}, but without considering a direct path or user mobility.

In this paper, we consider the estimation and tracking of the LOS channel between the UE and RIS, particularly considering large-scale mobility where there is no stationary channel statistics (contrary to previous works). We derive a new MLE for joint estimation of the UE-RIS and direct channel, with arbitrary RIS configurations during pilot transmission. We propose a mechanism where the estimation accuracy is progressively refined by updating the RIS configurations during pilot transmission, thereby enabling much fewer pilots than in \cite{Jensen2020a,Wang2020a,Papazafeiropoulos2022a,Zhang2022a,Xu2022,Chen2022a}. Finally, we demonstrate how the mechanism can be used for tracking the channel under user mobility.

\section{System Model}
\label{sec:sysmod}

An RIS-assisted communication system is considered, where the transmission between the single-antenna UEs and
the single-antenna BS is aided by an RIS with $N$ reconfigurable elements.
The channel between the BS and the RIS, $\vect{h} = [h_1,\ldots,h_N]^{\Ttran} \in \mathbb{C}^N$, is assumed to be known since the RIS and BS are deployed at fixed locations. The channel between an arbitrary UE and the RIS is denoted by $\vect{g} = [g_1,\ldots,g_N]^{\Ttran}  \in \mathbb{C}^N$ and the direct channel between this UE and the BS is represented by $d \in \mathbb{C}$. The BS serves a multitude of UEs at different locations in a time-division multiple access (TDMA) manner. The channels $\vect{g}$ and $d$ are time-varying in a non-stationary manner and must thus be estimated in the absence of a statistical characterization.
We consider the uplink but the results of this paper are also applicable to the downlink.

If the UE transmits the signal $x \in \mathbb{C}$ to the BS, the received signal can be expressed as \cite{Bjornson2022a}
\begin{equation}
\label{eq:received_signal}
 y = \left( \sum_{n=1}^{N} h_n g_n e^{-j\theta_n} + d \right) x + w, 
\end{equation}
where $w \sim \CN(0,\sigma^2 )$ is the complex Gaussian receiver noise and $\theta_n$ is the phase shift that the $n$th RIS element applies to the impinging signal. 
If $x_{\mathrm{d}} \sim \CN(0,P_{\mathrm{d}})$ is the transmitted data signal with power $P_{\mathrm{d}}$, the spectral efficiency (SE) for a given RIS configuration is \cite{Bjornson2022a}
\begin{align} \label{eq:rate-expression}
    &\log_2 \!\left( 1 + \left| \sum_{n=1}^{N} h_n g_n e^{-j\theta_n} + d \right|^2 \frac{P_{\mathrm{d}} }{\sigma^2} \right) \\ &\leq \log_2 \!\left( 1 + \left( \sum_{n=1}^{N} |h_n g_n| + |d| \right)^2 \!\frac{P_{\mathrm{d}}}{\sigma^2} \right), \label{eq:SE-expression}
\end{align}
where the expression on the first line holds for the arbitrary RIS phase shifts $e^{-j\theta_1},\ldots,e^{-j\theta_N}$ and the upper bound on the second line is achieved by setting the RIS phase shifts as $\theta_n = \arg(h_n)+\arg(g_n) - \arg(d)$ for $n=1,\ldots,N$. 
With this optimal selection, the signals from all the $N+1$ propagation paths are combined coherently at the BS. As $\vect{h}$ is assumed to be known, we need to estimate the direct channel $d$ and the channel $\vect{g}$ between the UE and RIS. To focus on the channel estimation problem, we consider a hardware implementation where any phase shifts $\theta_n \in [0,2\pi)$ can be selected.

In the next section, we will develop a novel MLE for joint estimation of $\vect{g}$ and  $d$, where $\vect{g}$ is assumed to be LOS-dominant such that  
\begin{equation}
\label{eq:A_set}
    \vect{g} \in \mathcal{A} = \left\{ \sqrt{\beta}e^{j\omega} \, \vect{a} ({\boldsymbol{\varphi}}): \beta \geq 0,\omega \in [0,2\pi), \boldsymbol{\varphi} \in \Phi \right\},
\end{equation}with $\beta$ and $\omega$ denoting the channel gain and the phase shift at the reference RIS element (first element), and $\vect{a}(\boldsymbol{\varphi})$ being the array response vector for a plane wave arriving from the angle-of-arrival (AOA) $\boldsymbol{\varphi}$.
Considering a planar RIS, we have $\boldsymbol{\varphi} = [\varphi_{\mathrm{az}}, \varphi_{\mathrm{el}}]^{\Ttran}$ where $\varphi_{\mathrm{az}}$ and $\varphi_{\mathrm{el}}$ are the azimuth and elevation AoAs and the set $\Phi$ is given by 

\begin{equation} 
\label{eq:Phi_set}
\Phi = \left\{\boldsymbol{\varphi} : \varphi_{\mathrm{az}} \in \left[- \frac{\pi}{2},\frac{\pi}{2} \right], \varphi_{\mathrm{el}} \in \left[- \frac{\pi}{2},\frac{\pi}{2} \right] \right\}.
\end{equation}The array response vector for the considered planar RIS can be expressed as 

\begin{align}
  &\vect{a}(\boldsymbol{\varphi}) = \Big [1,\ldots,e^{j(N_{\mathrm{H}} -1)\psi_{\mathrm{H}}},\ldots, \notag \\ &e^{j \left( (n_{\mathrm{H}} -1) \psi_{\mathrm{H}} + (n_{\mathrm{V}} - 1) \psi_{\mathrm{V}}\right)}, \ldots,e^{j\left((N_{\mathrm{H}} -1)\psi_{\mathrm{H}} + (N_{\mathrm{V}} -1)\psi_{\mathrm{V}}\right)}\Big]^{\Htran},
\end{align}where $n_{\mathrm{H}}$ and $n_{\mathrm{V}}$ are the row and column indices of the RIS elements, respectively, and $N_{\mathrm{H}}$ and $N_{\mathrm{V}}$ denote the number of elements in each row and column of the RIS such that $N = N_{\mathrm{H}} N_{\mathrm{V}}$. Further, 
\begin{equation}
\psi_{\mathrm{H}} = \frac{2\pi}{\lambda} \Delta_{\mathrm{H}} \sin (\varphi_{\mathrm{az}}) \cos(\varphi_{\mathrm{el}}),~~
\psi_{\mathrm{V}} = \frac{2\pi}{\lambda} \Delta_{\mathrm{V}} \sin (\varphi_{\mathrm{el}}), 
\end{equation}
with $\Delta_{\mathrm{H}}$ and $\Delta_{\mathrm{V}}$ indicating the inter-element spacing between the adjacent horizontal and vertical elements.

\section{Parametric Maximum Likelihood Estimator}

In this section, we will derive the proposed MLE for the problem at hand.
By defining the RIS phase-shift vector $\boldsymbol{\theta} = \left[e^{-j\theta_1},\ldots,e^{-j\theta_N}\right]^{\Ttran} \in \mathbb{C}^N$ and the diagonal matrix  $\vect{D}_{\vect{h}}= \diag (h_1,\ldots,h_N)$, we can rewrite \eqref{eq:received_signal} as 

\begin{equation}
   y = \left(\boldsymbol{\theta}^{\Ttran} \vect{D}_{\vect{h}} \vect{g} + d \right) x + w.
\end{equation}
The cascaded UE-RIS-BS  channel $\boldsymbol{\theta}^{\Ttran} \vect{D}_{\vect{h}} \vect{g}$ can be interpreted as a projection of the unknown channel $\vect{g}$ onto the direction of the vector $\boldsymbol{\theta}^{\Ttran} \vect{D}_{\vect{h}}$ determined by the RIS configuration. The UE transmits a known pilot signal at $L$ time instances and the RIS uses a different configuration at each time instance so that the projections of $\vect{g}$ in different directions are observable. Particularly, if the UE transmits the deterministic pilot signal $x_{\mathrm{p}} = \sqrt{P_\mathrm{p}}$ with power $P_{\mathrm{p}}$ at $L$ time instances with the RIS being configured as $\boldsymbol{\theta}_1,\ldots,\boldsymbol{\theta}_L$, the concatenated received signal $\vect{y} \in \mathbb{C}^L$ at the BS can be expressed as 
\begin{equation} \label{eq:received-pilot}
    \vect{y} = \left(\vect{B} \vect{D}_{\vect{h}} \vect{g} + d \vect{1}_{L \times 1} \right)\sqrt{P_\mathrm{p}} + \vect{w}
\end{equation} 
where
\begin{align}
   \label{eq:configuration_matrix} \vect{B} &= \left[\boldsymbol{\theta}_1,\ldots,\boldsymbol{\theta}_L\right]^{\Ttran}, \\
    \vect{w} &= [w_1,\ldots,w_L]^{\Ttran},
\end{align} 
and $w_l \sim \CN(0,\sigma^2 )$ is the independent noise at pilot time instance $l$, for  $l=1,\ldots,L$.

There is a multitude of channel estimators that can be developed based on the received pilot signal in \eqref{eq:received-pilot}.
We will take the MLE approach \cite{Kay1993a} because the channels $\vect{g}$ and $d$ are assumed to be unknown but not having a known random characterization due to non-stationary mobility. 
The probability density function (PDF) of $\vect{y}$ for a given $\vect{g}$ and $d$ can be expressed as 
\begin{equation}
\label{eq:y_PDF}
    f_{\vect{Y}}(\vect{y}; \vect{g},d) = \frac{1}{(\pi \sigma ^2)^L} e^{- \frac{\|\vect{y} - \left(\vect{B}\vect{D}_{\vect{h}} \vect{g} + d\vect{1} \right) \sqrt{P_\mathrm{p}}\|^2}{\sigma^2}}.
\end{equation} 
We are looking for the channel estimates $\hat{\vect{g}}$ and $\hat{\vect{d}}$ that maximize the PDF in \eqref{eq:y_PDF}, which is equivalent to minimizing the squared norm in the exponent. Hence, we can formulate the MLE problem as

\begin{align}
\label{eq:g_d_hat_general}
    &\{\hat{\vect{g}},\hat{d}\} =  \argmin{\vect{g} \in \mathbb{C}^N,d\in \mathbb{C}}~ \left\|\vect{y} - \left(\vect{B}\vect{D}_{\vect{h}} \vect{g} + d\vect{1} \right) \sqrt{P_\mathrm{p}} \right\|^2 \notag \\ & = \argmin{\vect{g} \in \mathbb{C}^N,d\in \mathbb{C}} P_\mathrm{p}\left\|\vect{B D_h g} + d\vect{1} \right\|^2 - 2\sqrt{P_\mathrm{p}} \mathrm{Re} \left(\vect{y}^{\Htran} (\vect{B D_h g} + d \vect{1})\right) \notag \\
    & = \argmin{\vect{g} \in \mathbb{C}^N,d\in \mathbb{C}}P_\mathrm{p} \left\| \vect{BD_h g} \right \|^2 - 2\sqrt{P_\mathrm{p}}\mathrm{Re} \left( \vect{y}^{\Htran} \vect{BD_h g}\right) + P_\mathrm{p} L |d|^2  \notag \\ & - 2\sqrt{P_\mathrm{p}}\mathrm{Re}\left ( d^* \vect{1}^{\Ttran} \left( \vect{y} - \sqrt{P_\mathrm{p}} \vect{BD_h g}\right)\right), 
\end{align}
where $\mathrm{Re}(\cdot)$ gives the real part of its argument.
Without loss of generality, we decompose the direct channel into its channel gain $\alpha \geq 0$ and phase shift $\vartheta \in [0,2\pi)$ as $d = \sqrt{\alpha} e^{j\vartheta}$. These channel parameters will only appear in the last two terms of \eqref{eq:g_d_hat_general}, thus the corresponding MLE subproblem can be expressed as
\begin{align}
   & \{\hat{\alpha},\hat{\vartheta}\} = \notag \\ &\argmin{\alpha \geq 0,\vartheta \in [0,2\pi) } P_\mathrm{p} L \alpha - 2\sqrt{P_\mathrm{p} \alpha}   \mathrm{Re}\left(e^{-j\vartheta} \vect{1}^{\Ttran} (\vect{y} - \sqrt{P_\mathrm{p}}\vect{BD_h g}) \right)\!.
\end{align} 
The minimum is found by first selecting the phase estimate
\begin{equation}
\label{eq:phase_d_hat1}
    \hat{\vartheta} = \arg \left( \vect{1}^{\Ttran} (\vect{y} - \sqrt{P_\mathrm{p}} \vect{BD_h g})\right),
\end{equation} 
which makes the term inside $\mathrm{Re}(\cdot)$ positive.
The resulting expression $P_{\mathrm{p}} L \alpha - 2\sqrt{P_{\mathrm{p}}} \sqrt{\alpha} |  \vect{1}^{\Ttran} (\vect{y} - \sqrt{P_{\mathrm{p}}}\vect{BD_h g}) |$ is a second-order polynomial of $\sqrt{\alpha}$ with the minimum given by
\begin{equation}
    \hat{\alpha} = \frac{\left| \vect{1}^{\Ttran} \left( \vect{y} - \sqrt{P_{\mathrm{p}}} \vect{BD_h g} \right)\right|^2}{P_{\mathrm{p}} L^2}.
\end{equation} 
By utilizing the parametrization of $\vect{g}$ given in \eqref{eq:A_set} and substituting $\hat{d} = \sqrt{\hat{\alpha}} e^{j\hat{\vartheta}}$ into \eqref{eq:g_d_hat_general}, the remaining MLE subproblem can be expressed as 
\begin{align}
\label{eq:channel_estimation}
    &\{\hat{\beta},\hat{\omega} ,\hat{\boldsymbol{\varphi}}\} = \notag \\ &\argmin{\substack{\beta \geq 0,\omega \in [0,2\pi),\\ \boldsymbol{\varphi} \in \Phi}} P_{\mathrm{p}}\beta  \left(\left \| \vect{BD_h a}(\boldsymbol{\varphi})\right\|^2  - \frac{1}{L}\left| \vect{1}^{\Ttran} \vect{BD_h a}(\boldsymbol{\varphi})\right|^2 \right)  \notag \\
    &-2\sqrt{P_{\mathrm{p}} \beta} \mathrm{Re}\left(e^{j\omega}\vect{y}^{\Htran}\left( \vect{I}_L - \frac{1}{L}\vect{1}_{L \times L} \right)\vect{BD_h}\vect{a}(\boldsymbol{\varphi})\right).
\end{align}
We first notice that $\omega$ only appears in the last term in \eqref{eq:channel_estimation}, which is minimized when the term inside $\mathrm{Re}(\cdot)$ is positive. The minimum is obtained by

\begin{align}
\label{eq:omega_estimate}\hat{\omega} &= - \arg \left(\vect{y}^{\Htran}\left( \vect{I}_L - \frac{1}{L}\vect{1}_{L \times L} \right)\vect{BD_h}\vect{a}(\boldsymbol{\varphi})\right). 
\end{align}Substituting \eqref{eq:omega_estimate} into \eqref{eq:channel_estimation} yields 
\begin{align}
\label{eq:channel_estimation2}
    \{\hat{\beta},\hat{\boldsymbol{\varphi}}\} &=\argmin{\beta \geq 0, \boldsymbol{\varphi} \in \Phi} P_{\mathrm{p}}\beta  \left(\left \| \vect{BD_h a}(\boldsymbol{\varphi})\right\|^2  - \frac{1}{L}\left| \vect{1}^{\Ttran} \vect{BD_h a}(\boldsymbol{\varphi})\right|^2 \right)  \notag \\
    &-2\sqrt{P_{\mathrm{p}} \beta} \left|\vect{y}^{\Htran}\left( \vect{I}_L - \frac{1}{L}\vect{1}_{L \times L} \right)\vect{BD_h}\vect{a}(\boldsymbol{\varphi})\right|,
\end{align}which is quadratic with respect to $\sqrt{\beta}$ and, thus, we have

\begin{align}
\label{eq:beta_estimate}
\hat{\beta} &= \frac{1}{P_{\mathrm{p}}} \left(\frac{\left| \vect{y}^{\Htran}\left( \vect{I}_L - \frac{1}{L}\vect{1}_{L \times L} \right)\vect{BD_h}\vect{a}(\boldsymbol{\varphi}) \right|}{\left \| \vect{BD_h a}(\boldsymbol{\varphi})\right\|^2  - \frac{1}{L}\left| \vect{1}^{\Ttran} \vect{BD_h a}(\boldsymbol{\varphi})\right|^2} \right) ^2. 
\end{align}

Finally, after substituting  \eqref{eq:beta_estimate} into \eqref{eq:channel_estimation2}, the MLE for the AoA is obtained as 
\begin{align}
\label{eq:AoA_estimate}
    \hat{\boldsymbol{\varphi}} &= \argmin{\boldsymbol{\varphi} \in \Phi} - \frac{\left| \vect{y}^{\Htran}\left( \vect{I}_L - \frac{1}{L}\vect{1}_{L \times L} \right)\vect{BD_h}\vect{a}(\boldsymbol{\varphi}) \right|^2}{\left \| \vect{BD_h a}(\boldsymbol{\varphi})\right\|^2  - \frac{1}{L}\left| \vect{1}^{\Ttran} \vect{BD_h a}(\boldsymbol{\varphi})\right|^2} \notag \\ &= \argmax{\boldsymbol{\varphi} \in \Phi}  \frac{\left| \vect{y}^{\Htran}\left( \vect{I}_L - \frac{1}{L}\vect{1}_{L \times L} \right)\vect{BD_h}\vect{a}(\boldsymbol{\varphi}) \right|^2}{\left \| \vect{BD_h a}(\boldsymbol{\varphi})\right\|^2  - \frac{1}{L}\left| \vect{1}^{\Ttran} \vect{BD_h a}(\boldsymbol{\varphi})\right|^2}. 
\end{align}
MLE subproblems of the kind in \eqref{eq:AoA_estimate} may have many local maxima \cite{Krim1996a}. We might identify all of them \cite{Wang2021a} but in our case, a simple grid search is sufficient because we will develop an algorithm in Section \ref{sec:adaptive_config} that iteratively improves the utility function in \eqref{eq:AoA_estimate} by sending new pilots until the global peak value is easily distinguished from the local ones. We thus solve \eqref{eq:AoA_estimate} numerically by performing a two-dimensional search over the set of feasible AoAs. 

The following theorem summarizes the proposed channel estimation scheme.
\begin{theorem}
    \label{th:main-result}
  
   The MLE of $d$ and $\vect{g}$ based on the received signal in \eqref{eq:received-pilot} are given by $\hat{d} = \hat{\alpha} e^{j\hat{\vartheta}}$ 
 and $\hat{\vect{g}} = \sqrt{\hat{\beta}} e^{j\hat{\omega}}\vect{a}(\hat{\boldsymbol{\varphi}})$ where
\vspace{-0.5mm}
 \begin{align}
 \label{eq:AoA_estimate_2D}\hat{\boldsymbol{\varphi}} &= \argmax{\boldsymbol{\varphi} \in \Phi}  \frac{\left| \vect{y}^{\Htran}\left( \vect{I}_L - \frac{1}{L}\vect{1}_{L \times L} \right)\vect{BD_h}\vect{a}(\boldsymbol{\varphi}) \right|^2}{\left \| \vect{BD_h a}(\boldsymbol{\varphi})\right\|^2  - \frac{1}{L}\left| \vect{1}^{\Ttran} \vect{BD_h a}(\boldsymbol{\varphi})\right|^2},\\
 \label{eq:phase_g_estimate}\hat{\omega} &= - \arg \left(\vect{y}^{\Htran}\left( \vect{I}_L - \frac{1}{L}\vect{1}_{L \times L} \right)\vect{BD_h}\vect{a}(\hat{\boldsymbol{\varphi}})\right), \\
   \label{eq:amp_g_estimate} \hat{\beta} &= \frac{1}{P_{\mathrm{p}}} \left(\frac{\left| \vect{y}^{\Htran}\left( \vect{I}_L - \frac{1}{L}\vect{1}_{L \times L} \right)\vect{BD_h}\vect{a}(\hat{\boldsymbol{\varphi}}) \right|}{\left \| \vect{BD_h a}(\hat{\boldsymbol{\varphi}})\right\|^2  - \frac{1}{L}\left| \vect{1}^{\Ttran} \vect{BD_h a}(\hat{\boldsymbol{\varphi}})\right|^2} \right) ^2,\\    
     \label{eq:phase_d_estimate}\hat{\vartheta} &= \arg \left( \vect{1}^{\Ttran} (\vect{y} - \sqrt{P_\mathrm{p}} \vect{BD_h} \hat{\vect{g}})\right),  \\
    \label{eq:amp_d_estimate}\hat{\alpha} &= \frac{\left| \vect{1}^{\Ttran} \left( \vect{y} - \sqrt{P_p} \vect{BD_h}\hat{\vect{g}} \right)\right|^2}{P_{\mathrm{p}} L^2}.
 \end{align}
\end{theorem}

The estimates are interconnected in the sense that
we must first compute $\hat{\boldsymbol{\varphi}}$ using \eqref{eq:AoA_estimate_2D} and use it to compute the MLE of
$\hat{\vect{g}}$ from \eqref{eq:phase_g_estimate} and \eqref{eq:amp_g_estimate}. Finally, $\hat{\vect{g}}$ is used to compute $\hat{d}$ using \eqref{eq:phase_d_estimate} and \eqref{eq:amp_d_estimate}.
This is the opposite order of how they were derived. 
We can now configure the RIS based on the estimates. Since there is no channel statistics, the RIS should be configured as if the estimates are perfect. Particularly, the phase shift vector 
\begin{equation}
\label{eq:optimal_phases}
    \bar{\boldsymbol{\theta}} = e^{j (\hat{\vartheta} - \hat{\omega})}\diag \left(e^{-j\mathrm{arg}(h_1)},\ldots,e^{-j\mathrm{arg}(h_N)} \right) \vect{a}^*(\hat{\boldsymbol{\varphi}}), 
\end{equation}achieves the maximum SE in \eqref{eq:SE-expression} if the parameters have been perfectly estimated. We notice that it depends on the AoA $\hat{\boldsymbol{\varphi}}$ and phase shifts $\hat{\omega}, \hat{\vartheta}$, but not on the channel gains $\hat{\alpha},\hat{\beta}$ since we only want to align all signal paths in phase.

\section{Adaptive RIS Configuration During Estimation}
\label{sec:adaptive_config}

The accuracy of the proposed MLE procedure in Theorem~\ref{th:main-result} depends on how the RIS is configured during channel estimation; that is, the choice of $\vect{B}$. Many previous works (e.g., \cite{Jensen2020a,Wang2020a,Wang2021a}) require $L \geq N$ since $\vect{B}$ is either selected a priori to explore all channel dimensions of $\mathbb{C}^N$ or randomly. In this section, we propose an iterative algorithm to adaptively add columns to $\vect{B}$ to improve the MLE accuracy. If we use the subscript $l$ to denote the estimated parameters in the $l$th iteration, from \eqref{eq:optimal_phases}, the SE-maximizing RIS phase shift vector is given by  
\vspace{-1mm}
\begin{equation}
\label{eq:optimal_config}
    \bar{\boldsymbol{\theta}}_l = e^{j (\hat{\vartheta}_l - \hat{\omega}_l)}\diag \left(e^{-j\mathrm{arg}(h_1)},\ldots,e^{-j\mathrm{arg}(h_N)} \right) \vect{a}^*(\hat{\boldsymbol{\varphi}}_l). 
\end{equation}This RIS phase shift vector is then used for selecting the next RIS configuration during the pilot transmission. 

We define a set $\mathcal{B}= \{ \bar{\boldsymbol{\varphi}}_1,\bar{\boldsymbol{\varphi}}_2,\ldots,\bar{\boldsymbol{\varphi}}_N \}$ of $N = N_H N_V$ plausible azimuth-elevation AoA pairs for the user that might be considered during channel estimation. 
Based on these angles, we define the resulting set of RIS configurations as
\begin{align}
\label{eq:configuration_set}
 \Theta = \left\{\diag \left(e^{-j\mathrm{arg}(h_1)},\ldots,e^{-j\mathrm{arg}(h_N)} \right) \vect{a}^*(\boldsymbol{\varphi}) : \boldsymbol{\varphi} \in \mathcal{B} \right \}.  
\end{align}
\vspace{-1mm}
Our proposed method can be explained as follows: First, we select two RIS configurations $\boldsymbol{\theta}_1$ and $\boldsymbol{\theta}_2$ from the set in \eqref{eq:configuration_set} and transmit pilots at two time instances utilizing the selected RIS configuration matrix $\vect{B}_2 = [\boldsymbol{\theta}_1,\boldsymbol{\theta}_2]^{\Ttran}$. We then use Theorem \ref{th:main-result} to estimate the AoA vector from \eqref{eq:AoA_estimate_2D} via a 2D search over the set defined in \eqref{eq:Phi_set}  and obtain the estimated phase shifts from \eqref{eq:phase_g_estimate} and \eqref{eq:phase_d_estimate}. For the $(l+1)$th pilot transmission, the RIS configuration is set as the unused configuration in $\Theta$ which is closest to the RIS phase shift vector $\bar{\boldsymbol{\theta}}_l$ in \eqref{eq:optimal_config}, in terms of the absolute value of the inner product. The RIS configuration matrix is updated as $\vect{B}_{l+1} = [\vect{B}_l^{\Ttran}, \boldsymbol{\theta}_{l+1}]^{\Ttran}$, where the subscripts indicate the iteration number. A new pilot is then transmitted using the  newly obtained configuration and the concatenated received signal $\vect{y}_{l+1} = [\vect{y}_{l}^{\Ttran},y_{l+1}]^{\Ttran}$ is used for channel estimation in the $(l+1)$th iteration. This procedure is iteratively performed until we reach the intended number of pilot transmissions $L$. Algorithm \ref{Alg:ML_Estimation} summarizes the proposed iterative scheme for estimating the channel parameters.

\begin{algorithm}[h]
\small
\caption{Parametric MLE of $\vect{g}$ and $d$.}
\label{Alg:ML_Estimation}
\begin{algorithmic}[1]
\STATE{Define the set $\Theta$ of plausible RIS configurations in \eqref{eq:configuration_set}}
\STATE{Select two initial RIS configurations $\boldsymbol{\theta}_1$ and $\boldsymbol{\theta}_2$ from $\Theta$}
\STATE{Set $\vect{B}_2 = [\boldsymbol{\theta}_1,\boldsymbol{\theta}_2]^{\Ttran}$ and update $\Theta \gets \Theta \setminus \{\boldsymbol{\theta}_1,\boldsymbol{\theta}_2\}$}
\STATE{Send pilot signals using the RIS configurations $\boldsymbol{\theta}_1,\boldsymbol{\theta}_2$ to get the received signal $\vect{y}_2 \in \mathbb{C}^2$}
 \FOR{$l = 2,\ldots,L$}
    \STATE{Compute $\hat{\boldsymbol{\varphi}}_l  =\argmax{\boldsymbol{\varphi} \in \Phi}~ \frac{\left| \vect{y}^{\Htran}\left( \vect{I}_l - \frac{1}{l}\vect{1}_{l \times l} \right)\vect{B}_l\vect{D_h}\vect{a}(\boldsymbol{\varphi}) \right|^2}{\left \| \vect{B}_l\vect{D_h a}(\boldsymbol{\varphi})\right\|^2  - \frac{1}{l}\left| \vect{1}^{\Ttran} \vect{B}_l\vect{D_h a}(\boldsymbol{\varphi})\right|^2}$}
    \STATE {Obtain $\hat{\omega}_l$, $\hat{\beta}_l$, and $\hat{\vartheta}_l$ from \eqref{eq:phase_g_estimate}-\eqref{eq:phase_d_estimate}}
    \IF{$l < L$}
    \STATE{Compute $\bar{\boldsymbol{\theta}}_l$ in \eqref{eq:optimal_config}}
    \STATE{Compute $\boldsymbol{\theta}_{l+1} = \argmax{\boldsymbol{\theta} \in \Theta}~ | \bar{\boldsymbol{\theta}}_l^{\Htran} \boldsymbol{\theta}|$}
    \STATE{Set $\vect{B}_{l+1} = [\vect{B}_l^{\Ttran}, \boldsymbol{\theta}_{l+1}]^{\Ttran}$, update $\Theta \gets \Theta \setminus \{\boldsymbol{\theta}_{l+1}\}$}
    \STATE{Send a pilot signal using the RIS configuration $\boldsymbol{\theta}_{l+1}$ and collect received signals in $\vect{y}_{l+1} = [\vect{y}_{l}^{\Ttran},y_{l+1}]^{\Ttran}$}
\ENDIF
 \ENDFOR 
 \STATE {Obtain $\hat{\alpha}$ \eqref{eq:amp_d_estimate}}
\RETURN  $\hat{\vect{g}} = \sqrt{\hat{\beta}_L}e^{j\hat{\omega}_L}\vect{a}(\hat{\boldsymbol{\varphi}}_L)$ and $\hat{d} = \sqrt{\hat{\alpha}} e^{j\hat{\vartheta}_L}$
 \end{algorithmic}
\end{algorithm}

We show in Section~\ref{sec:results} that the proposed algorithm quickly converges to a good channel estimate using $L\ll N$.

\subsection{Channel Tracking}

If the BS has a sense of where the UE is, the initial RIS configurations $\boldsymbol{\theta}_1,\boldsymbol{\theta}_2$ in the channel estimation can be selected to explore that location.
Hence, one important use case of Algorithm \ref{Alg:ML_Estimation} is to successively estimate the channel of a UE that moves along an unknown trajectory. If $\boldsymbol{\theta}_1$ is selected based on the last successful RIS configuration, we begin searching for a new RIS configuration in its vicinity. This can lead to efficient tracking of the RIS channel using even fewer pilots.
We will detail such a procedure in the next section.

\section{Numerical Results}
\label{sec:results}

We will now demonstrate the effectiveness of the proposed parametric MLE and adaptive RIS configuration. We consider an indoor setup where the RIS is mounted in the middle of a wall in a $5 \times 5~\mathrm{m}^2$  room, to have a LOS link to a UE in the room and to the BS outside the building (e.g., through a window). The RIS is equipped with $N_\mathrm{H} = N_\mathrm{V} = 8$ elements where the element spacing is $\Delta_\mathrm{H} = \Delta_\mathrm{V} = \lambda/4$. Since the BS is deployed outside the building, the direct path between UE and BS is a non-LOS link. We assume that the direct path is ten times stronger than the per-element cascaded channel and generated as $d \sim \mathcal{N}_{\mathbb{C}}(0,10|h_n g_n|^2)$. Since we consider LOS channels to/from the RIS, $|h_n g_n|$ is independent of the element index $n$. We generate the azimuth and elevation angles in the set $ \mathcal{B}$ considered during channel estimation as \cite{Bjornson2022b}: 
\begin{align}
    &\!\varphi_\mathrm{az} \!= \arcsin \left(\frac{2m}{N_\mathrm{H}} \right) : m = -\left\lfloor \frac{N_\mathrm{H}-1}{2} \right\rfloor \!, \cdots, \left\lfloor \frac{N_\mathrm{H}}{2} \right\rfloor  \\
    &\!\varphi_\mathrm{el} \!= \arcsin \left( \frac{2m}{N_\mathrm{V}} \right) : m = -\left\lfloor \frac{N_\mathrm{V}-1}{2} \right\rfloor \!, \cdots, \left\lfloor \frac{N_\mathrm{V}}{2} \right\rfloor.
\end{align}

\begin{figure}[t!]
        \centering
        \begin{subfigure}[b]{\columnwidth} \centering
	\begin{overpic}[width=0.9\columnwidth,tics=10]{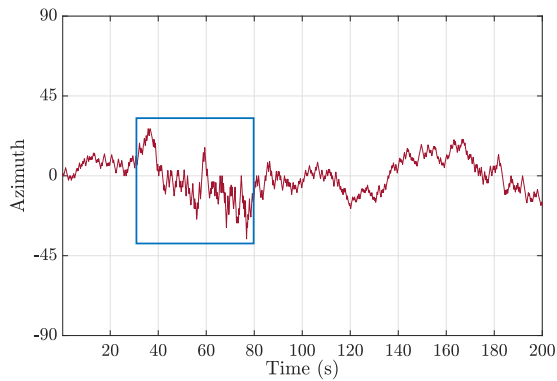}
\end{overpic}  \vspace{-3mm}
                \caption{Variations in the azimuth angle over time.}
    \vspace{+2mm}
        \end{subfigure} 
        \begin{subfigure}[b]{0.9\columnwidth} \centering
	\begin{overpic}[width=\columnwidth,tics=10]{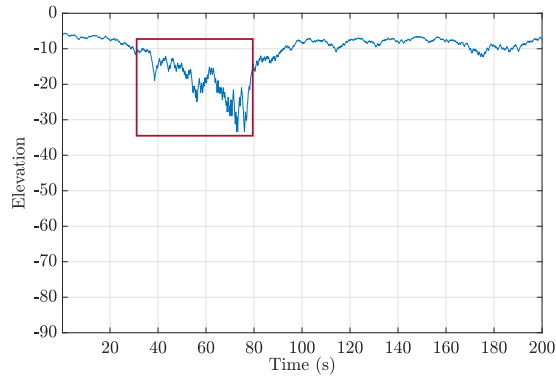}
\end{overpic} \vspace{-3mm}
                \caption{Variations in the elevation angle over time.} 
        \end{subfigure} 
        \caption{An example of how the azimuth and elevation change over time when a UE walks randomly in the considered room.} 
        \label{fig:AzElEvolution}  
\end{figure}

The user walks randomly in the room with an average speed of $50 \, \mathrm{cm}/\mathrm{s}$ such that the azimuth and elevation angles seen from the RIS change over time. An instance of the angle variation is illustrated in Fig.~\ref{fig:AzElEvolution} for $200$ seconds. When the UE is close to the RIS, a relatively small move may translate to a rapid variation in the azimuth and elevation angles. In Fig.~\ref{fig:AzElEvolution}, this phenomenon occurred in the time interval indicated by a rectangle. We utilize these angles to generate channel vector $\mathbf{g}$ as in \eqref{eq:A_set}. Since the LOS channel between the RIS and BS is known, our algorithm compensates for $\vect{D}_{\vect{h}}$ by designing the RIS configuration according to \eqref{eq:configuration_set}. Therefore, any choice of the angles used to generate $\mathbf{h}$ gives the same results.

We define the per-element data SNR as 
\begin{equation}
   \mathrm{SNR}_d = \frac{P_{\mathrm{d}}}{\sigma^2} | h_n g_n |^2,
\end{equation}
and assume the pilot SNR, $\mathrm{SNR}_{\mathrm{p}}$, is $10$\,dB larger than $\mathrm{SNR}_{\mathrm{d}}$ (i.e., $P_{\mathrm{p}} = 10 P_{\mathrm{d}}$) since pilots can be expanded over frequency. 

\begin{figure}[t]
    \centering
    \includegraphics[width= \linewidth]{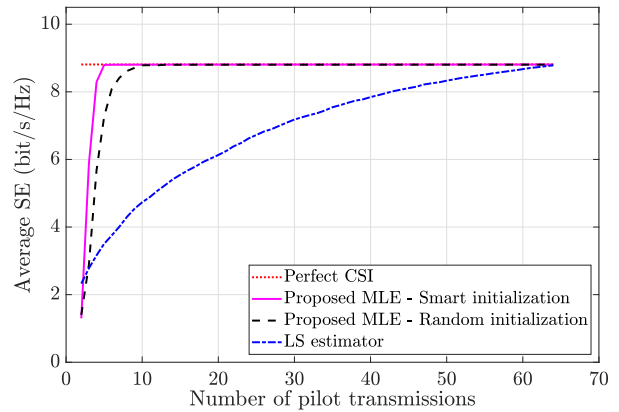}
    \caption{Average SE with respect to pilot length $L$. The proposed method is compared with with LS and perfect CSI.}
    \label{fig:SNRVsPilot}
\end{figure}

Fig.~\ref{fig:SNRVsPilot} shows the average SE over different noise realization and all AOAs during the random walk, with respect to the pilot length $L$ when $\mathrm{SNR}_{\mathrm{p}} = 0$ dB and $\mathrm{SNR}_{\mathrm{d}} = -10$ dB. In this figure, the black and magenta lines represent the SE in \eqref{eq:rate-expression} achieved using the proposed MLE scheme, where the colors distinguish between different approaches of selecting $\{\boldsymbol{\theta}_1,\boldsymbol{\theta}_2\}$ in Step $2$ of Algorithm~\ref{Alg:ML_Estimation}. The dashed black line corresponds to the random selection of $\boldsymbol{\theta}_1$ and $\boldsymbol{\theta}_2$ from the set $\Theta$ while the solid magenta line follows a smarter selection. Specifically, when a user wishes to connect to the BS for the first time, Algorithm~\ref{Alg:ML_Estimation} selects $\{\boldsymbol{\theta}_1,\boldsymbol{\theta}_2\}$ randomly which results in the SE represented with the black line in Fig.~\ref{fig:SNRVsPilot}. Henceforth, the CSI only needs to be updated periodically to avoid SE losses. As the user speed is limited, we select $\boldsymbol{\theta}_1$ based on our last successful RIS configuration. Specifically, if $\bar{\boldsymbol{\theta}}_L$ is the RIS phase shift vector obtained in the last channel estimation phase, we select $\boldsymbol{\theta}_1$ as
\begin{equation}
\label{eq:smartInit}
    \boldsymbol{\theta}_1 = \argmax{\boldsymbol{\theta} \in \Theta}~ | \bar{\boldsymbol{\theta}}_L^{\Htran} \boldsymbol{\theta}|,
\end{equation}
and $\boldsymbol{\theta}_2$ randomly. This approach results in the magenta line in Fig.~\ref{fig:SNRVsPilot}. We compare our method with the maximum SE obtained with perfect CSI according to \eqref{eq:SE-expression}. Additionally, we use the conventional LS estimator \cite{Jensen2020a} as the benchmark, where $L$ columns of an $N \times N$ DFT matrix are used as the $L$ RIS configurations \cite{Bjornson2022b}. We observe that the black line converges gradually to the maximum SE and reaches $98\%$ with $L = 9$ pilot transmissions. When we exploit our last RIS configuration to initialize the channel estimation algorithm at the next time instance, $98\%$ of the maximum 
 SE is achieved already with $L = 5$ pilots. On the other hand, the LS estimator needs $60$ pilots to attain this level. We stress that our proposed algorithm converges quickly since, at each iteration, it sends a new pilot using an unused RIS configuration from $\Theta$ that helps solidify estimation accuracy.

\begin{figure*}[!ht]
    \centering
    \includegraphics[width = \linewidth]{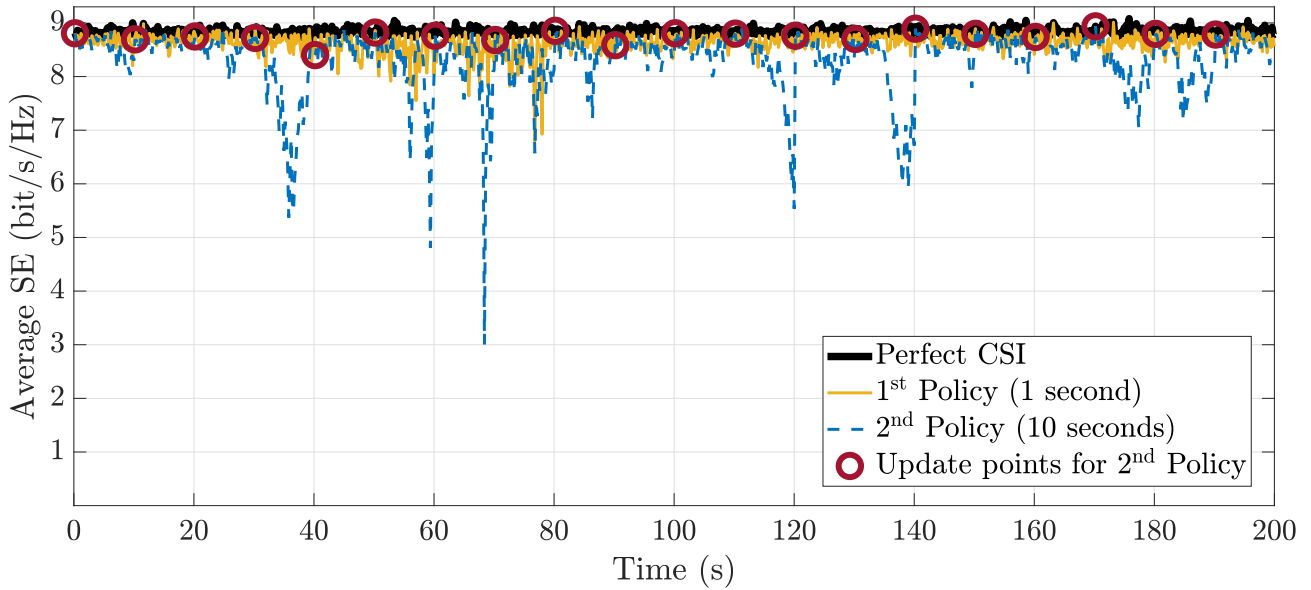}
    \caption{Achievable SE following two different policies to reconfigure the RIS. The channel instances are generated every $200\,$ms and the UE moves $10\,$cm between each two instances. $L = 6$ pilots are used to update the CSI and re-configure the RIS.}
    \label{fig:UpdateFreq}
\end{figure*}

\subsection{Channel Tracking}

The optimized RIS configuration in \eqref{eq:optimal_phases} beamforms the impinging signal toward the UE and matches the phase shift with that of the direct path. When the user moves randomly in the room, the AOAs at the RIS and the phase difference between the direct and cascaded path $(\vartheta - \omega)$ will change. The variations have therefore non-stationary statistics and are related to the walking speed of the user and its current distance from the RIS. 
We will demonstrate the effectiveness of the proposed algorithm to track the channel variations and investigate how frequently the RIS configuration must be changed to maintain an acceptable SE. 
To this end, we generate channel realizations every $200\,$ms and between two time instances, the UE moves $10 \,$cm. To re-estimate the channel and  re-configure the RIS, we use Algorithm~\ref{Alg:ML_Estimation} with $L = 6$ and follow the smart initialization given in \eqref{eq:smartInit} that resulted in the magenta line in Fig.~\ref{fig:SNRVsPilot}. Accordingly, Fig.~\ref{fig:UpdateFreq} reports the SE (averaged over different noise realizations) at each time instance, based on two RIS re-configuration policies. The first policy estimates the channel and selects the RIS configuration every $1$ second, while the second policy updates the configuration every $10$ seconds. For both policies, the RIS configuration is constant between two updates. We observe that with the first policy, we can track the channel variations and maintain an SE close to the perfect CSI case almost all the time; the solid yellow line rarely falls below $8$ bps/Hz, which is $92\%$ of the maximum SE in this setup.
On the other hand, updating the CSI every $10$ seconds is not frequent enough to efficiently track the channel, leading to many deep falls in SE (see the blue dashed curve) since the UE moves out of the main lobe of the beam from the RIS. The intuition is that as long as the RIS is reconfigured frequently enough so that the UE never moves out of the main lobe, we have an insignificant performance loss in between the RIS updates which is mainly due to the fast-changing phase shift difference between the direct and RIS paths.

\section{Conclusions}

In this paper, we proposed a new MLE-based channel estimation framework with low training overhead, high accuracy, and mobility management capability. By adaptively configuring the RIS during the channel estimation process, the proposed method progressively improves the channel estimation accuracy and obtains accurate CSI using much fewer pilots than there are RIS elements. We showed that the number of pilot transmissions can be further reduced if we have some prior knowledge of the location of the user. This feature can be used to efficiently track the channel of a user that moves along an unknown trajectory with very low pilot overhead.

\bibliographystyle{IEEEtran}
\bibliography{refs}

\begin{thebibliography}{10}
\providecommand{\url}[1]{#1}
\csname url@samestyle\endcsname
\providecommand{\newblock}{\relax}
\providecommand{\bibinfo}[2]{#2}
\providecommand{\BIBentrySTDinterwordspacing}{\spaceskip=0pt\relax}
\providecommand{\BIBentryALTinterwordstretchfactor}{4}
\providecommand{\BIBentryALTinterwordspacing}{\spaceskip=\fontdimen2\font plus
\BIBentryALTinterwordstretchfactor\fontdimen3\font minus
  \fontdimen4\font\relax}
\providecommand{\BIBforeignlanguage}[2]{{%
\expandafter\ifx\csname l@#1\endcsname\relax
\typeout{** WARNING: IEEEtran.bst: No hyphenation pattern has been}%
\typeout{** loaded for the language `#1'. Using the pattern for}%
\typeout{** the default language instead.}%
\else
\language=\csname l@#1\endcsname
\fi
#2}}
\providecommand{\BIBdecl}{\relax}
\BIBdecl

\bibitem{Huang2018a}
C.~Huang, A.~Zappone, G.~C. Alexandropoulos, M.~Debbah, and C.~Yuen,
  ``Reconfigurable intelligent surfaces for energy efficiency in wireless
  communication,'' \emph{IEEE Trans. Wireless Commun.}, vol.~18, no.~8, pp.
  4157--4170, 2019.

\bibitem{Renzo2020b}
M.~D. Renzo \emph{et~al.}, ``Smart radio environments empowered by
  reconfigurable intelligent surfaces: How it works, state of research, and
  road ahead,'' \emph{IEEE J. Sel. Areas Commun.}, vol.~38, no.~11, pp.
  2450--2525, 2020.

\bibitem{Bjornson2020a}
E.~Bj\"ornson, {\"O}.~\"Ozdogan, and E.~G. Larsson, ``Reconfigurable
  intelligent surfaces: Three myths and two critical questions,'' \emph{IEEE
  Commun. Mag.}, no.~12, pp. 90--96, 2020.

\bibitem{Bjornson2022a}
E.~Bj\"{o}rnson, H.~Wymeersch, B.~Matthiesen, P.~Popovski, L.~Sanguinetti, and
  E.~de~Carvalho, ``Reconfigurable intelligent surfaces: A signal processing
  perspective with wireless applications,'' \emph{IEEE Signal Process. Mag.},
  vol.~39, no.~2, pp. 135--158, 2022.

\bibitem{Jensen2020a}
T.~L. {Jensen} and E.~{De Carvalho}, ``An optimal channel estimation scheme for
  intelligent reflecting surfaces based on a minimum variance unbiased
  estimator,'' in \emph{IEEE ICASSP}, 2020, pp. 5000--5004.

\bibitem{Wang2020a}
Z.~Wang, L.~Liu, and S.~Cui, ``Channel estimation for intelligent reflecting
  surface assisted multiuser communications: Framework, algorithms, and
  analysis,'' \emph{IEEE Trans. Wireless Commun.}, vol.~19, no.~10, pp.
  6607--6620, 2020.

\bibitem{Papazafeiropoulos2022a}
A.~Papazafeiropoulos, I.~Krikidis, and P.~Kourtessis, ``Impact of channel aging
  on reconfigurable intelligent surface aided massive {MIMO} systems with
  statistical {CSI},'' \emph{IEEE Trans. Veh. Technol.}, pp. 1--15, 2022.

\bibitem{Zhang2022a}
Y.~Zhang, J.~Zhang, D.~W.~K. Ng, H.~Xiao, and B.~Ai, ``Performance analysis of
  reconfigurable intelligent surface assisted systems under channel aging,''
  \emph{Intelligent and Converged Networks}, vol.~3, no.~1, pp. 74--85, 2022.

\bibitem{Xu2022}
C.~Xu, J.~An, T.~Bai, S.~Sugiura, R.~G. Maunder, Z.~Wang, L.-L. Yang, and
  L.~Hanzo, ``Channel estimation for reconfigurable intelligent surface
  assisted high-mobility wireless systems,'' \emph{IEEE Trans. Veh. Technol.},
  pp. 1--16, 2022.

\bibitem{Chen2022a}
Y.~Chen, Y.~Wang, and L.~Jiao, ``Robust transmission for reconfigurable
  intelligent surface aided millimeter wave vehicular communications with
  statistical csi,'' \emph{IEEE Trans. Wireless Commun.}, vol.~21, no.~2, pp.
  928--944, 2022.

\bibitem{CYou2020}
C.~You, B.~Zheng, and R.~Zhang, ``Channel estimation and passive beamforming
  for intelligent reflecting surface: Discrete phase shift and progressive
  refinement,'' \emph{IEEE J. Sel. Areas Commun.}, vol.~38, no.~11, pp.
  2604--2620, Nov. 2020.

\bibitem{Swindlehurst2022a}
A.~L. Swindlehurst, G.~Zhou, R.~Liu, C.~Pan, and M.~Li, ``Channel estimation
  with reconfigurable intelligent surfaces--a general framework,'' \emph{Proc.
  IEEE}, vol. 110, no.~9, pp. 1312--1338, 2022.

\bibitem{Wang2021a}
W.~Wang and W.~Zhang, ``Joint beam training and positioning for intelligent
  reflecting surfaces assisted millimeter wave communications,'' \emph{IEEE
  Trans. Wireless Commun.}, vol.~20, no.~10, pp. 6282--6297, 2021.

\bibitem{Bjornson2022b}
E.~Bj\"ornson and P.~Ramezani, ``Maximum likelihood channel estimation for
  {RIS}-aided communications with {LOS} channels,'' in \emph{Asilomar
  Conference on Signals, Systems and Computers}, 2022.

\bibitem{Kay1993a}
S.~M. Kay, \emph{Fundamentals of statistical signal processing: Estimation
  theory}.\hskip 1em plus 0.5em minus 0.4em\relax Prentice Hall, 1993.

\bibitem{Krim1996a}
H.~Krim and M.~Viberg, ``Two decades of array signal processing research: the
  parametric approach,'' \emph{IEEE Signal Process. Mag.}, vol.~13, no.~4, pp.
  67--94, 1996.

\end{thebibliography}

\end{document}